# Could we identify hot Ocean-Planets with CoRoT, *Kepler* and Doppler velocimetry?

--------

F. Selsis [1], B. Chazelas [2], P. Bordé [3,*], M. Ollivier [2], F. Brachet [2], M. Decaudin [2], F. Bouchy [4], D. Ehrenreich [4], J.-M. Grießmeier [5], H. Lammer [6], C. Sotin [7], O. Grasset [7], C. Moutou [8], P. Barge [8], M. Deleuil [8], D. Mawet [9], D. Despois [10], J. F. Kasting [11], A. Léger [2]

[1] Ecole Normale Supérieure de Lyon, Centre de Recherche Astronomique de Lyon, 46 allée d'Italie, F-69364 Lyon Cedex 07, France ; CNRS, UMR 5574 ; Université de Lyon 1, Lyon, France., franck.selsis@ens-lyon.fr

[2] Institut d'Astrophysique. Spatiale, bat 121, Université Paris–Sud and CNRS (UMR 8617); Univ. Paris-Sud, F-91405 Orsay; Fr, bruno.chazelas@ias.u-psud.fr; Marc.Ollivier@ias.u-psud.fr; Frank.Brachet@ias.u-psud.fr; michel.decaudin@ias.u-psud.fr; Alain.Leger@ias.u-psud.fr

[3] Harvard-Smithsonian Center for Astrophysics, 60 Garden Street, Cambridge, MA 02138, USA, pborde@ipac.caltech.edu

[*] Michelson Postdoctoral Fellow. Current address: Michelson Science Center, California Institute of Technology, 770 S Wilson Avenue, MS 100-22, Pasadena, CA 91125, USA

[4] Institut d'Astrophysique de Paris ; CNRS (UMR 7095) ; Université Pierre & Marie Curie ; 98, bis boulevard Arago, F-75014 Paris, France ; bouchy@iap.fr; ehrenrei@iap.fr;

[5] LESIA, CNRS-Observatoire de Paris, 92195 Meudon, France, jean-mathias.griessmeier@obspm.fr

[6] Space Research Institute, Austrian Academy of Sciences, Schmiedlstr. 6, A-8042, Graz, Austria, helmut.lammer@oeaw.ac.at

[7] Géophysique, Université de Nantes, F-44321 Nantes cedex 3, Fr, sotin@chimie.univ-nantes.fr; Olivier.Grasset@chimie.univ-nantes.fr

[8] Laboratoire d'Astrophysique de Marseille (LAM/OAMP), CNRS, – BP 8 - Traverse du Siphon, 13376 Marseille Cedex 12, Claire.Moutou@oamp.fr; pierre.barge@oamp.fr; magali.deleuil@oamp.fr

[9] université de Liège, 17 allée du 6 Août, 4000 Sart-Tilman, Belguim, dimitri.mawet@ulg.ac.be

[10] Observatoire de Bordeaux (INSU/CNRS), B.P. 89, F-33270 Floirac, Fr, Didier.Despois@obs.u-bordeaux1.fr

[11] Dept. of Geosciences, The Pennsylvania State University, University Park, Pennsylvania 16802, USA, kasting@geosc.psu.edu

ABSTRACT

Planets less massive than about 10 $M_{Earth}$ are expected to have no massive H-He atmosphere and a cometary composition (~ 50% rocks, 50% water, by mass) provided they formed beyond the snowline of protoplanetary disks. Due to inward migration, such planets could be found at any distance between their formation site and the star. If migration stops within the habitable zone, this may produce a new kind of planets, called *Ocean-Planets*. Ocean-planets typically consist in a silicate core, surrounded by a thick ice mantle, itself covered by a 100 km – deep ocean. The possible existence of ocean-planets raises important astrobiological questions: Can life originate on such body, in the absence of continent and ocean-silicate interfaces? What would be the nature of the atmosphere and the geochemical cycles ?

In this work, we address the fate of *Hot Ocean-Planets* produced when migration ends at a closer distance. In this case the liquid/gas interface can disappear, and the hot $H_2O$ envelope is made of a supercritical fluid. Although we do not expect these bodies to harbor life, their detection and identification as water-rich planets would give us insight as to the abundance of hot and, by extrapolation, cool Ocean-Planets.

The water reservoir of these planets seems to be weakly affected by gravitational escape, provided that they are located beyond some minimum distance, *e.g.* 0.04 AU for a 5-Earth-mass planet around a Sun-like star. The swelling of their water atmospheres by the high stellar flux is expected not to significantly increase the planets' radii. We have studied the possibility of detecting and characterizing these Hot Ocean-Planets by measuring their mean densities using transit missions in space – CoRoT (CNES) and *Kepler* (NASA) – in combination with Doppler velocimetry from the ground – HARPS (ESO) and possible future instruments. We have determined the domain in the [stellar magnitude, orbital distance] plane where discrimination between Ocean-Planets and rocky planets is possible with these instruments.

The brightest stars of the mission target lists and the planets closest to their stars are the most favorable cases. Full advantage of high precision photometry by CoRoT, and particularly *Kepler*, can be obtained only if a new generation of Doppler instruments is built.

## 1. Introduction

### 1.1. Water-rich vs silicate-rich planets

Ocean-planets (OPs) represent a new kind of planets, recently proposed by Léger et al. (2004, 2003) and Kuchner (2003), which nature can be described as follows:

(1) Their formation takes place beyond the snow line, giving them a cometary-like bulk composition: silicates and water, in roughly equal amounts by mass. Planets that accreted planetesimals formed at less than about 3 AU from a solar type star have a much lower water content ($w \equiv H_2O$/silicate, by mass): $w$ is $5 \times 10^{-4}$ for the Earth and remains below 10% for a planet entirely made of carbonaceous-chondritic material. Raymond et al. (2006b, 2007) simulated the formation of habitable planets in the absence of migration and found $w < 5\%$ at the end of the accretion. By simulating planetary formation after the migration of an outer giant planet to a close-in orbit, Raymond et al. (2006a) predicted that the mixing of material from the inner and outer regions can produce planets in the habitable zone consisting of up to ∼30% water. This might be another way to form OPs.

(2) Unlike Uranus and Neptune, OPs do not reach a sufficient mass to accrete a gaseous envelope directly from the protoplanetary disk. They are thus less massive than about 10 $M_{Earth}$ and assumed to have no $H_2$–He envelope. This assumption is robust for planets of 6 $M_{Earth}$ and below, and has to be discussed further for planets in the range 6–10 $M_{Earth}$ for which significant gas accretion may still have occurred (Alibert et al., 2006; Rafikov, 2006).

(3) To become an OP, the rocky–icy planet has to migrate to the inner part of the planetary system through type-I migration (Papaloizou et al., 2006). If migration stops in the habitable zone (HZ) of its star, the planet becomes what Léger et al. (2004, 2003) call an ocean-planet (OP). When migration stops at shorter orbital distances, it can give birth to planets with a thick and hot $H_2O$ envelope with no liquid–atmosphere interface.

(4) Planets of a few $M_{Earth}$ are expected to be mainly made of silicate and water, except when found around stars with a C/O ratio significantly higher than solar, where *carbon-planets* may form (Kuchner and Seager, 2005). Although one should keep in mind this possibility for some planetary systems, we consider in this paper that planets can be characterized with a single number: the water content $w$. In this case, internal structure models provide different planetary radii, for a given mass, as a function of this water ratio. For a 6-Earth-mass planet, Léger et al. (2004, 2003) found $R_{pl} = 2.00\,R_{Earth}$ for an OP and $R_{pl} = 1.63\,R_{Earth}$ for a rocky one. Sotin et al. (2007) generalized these calculations for masses in the (0.01–10 $M_{Earth}$) range and variable water content (Fig. 1). Independently, Valencia et al. (2006) also derived mass–radius curves for planets with various water contents. These two works are in very good agreement. Both found that, in all cases, a couple of values ($M_{pl}$, $R_{pl}$) are indicative of a planetary composition.

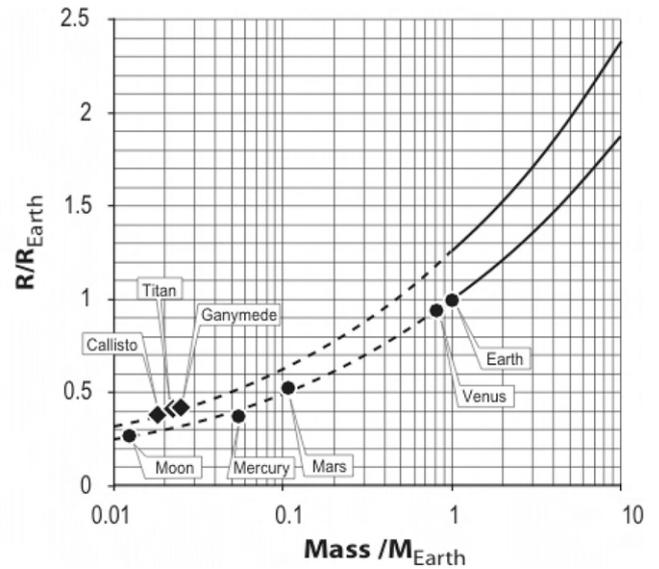

Fig. 1. Relation between the mass and the radius of ocean-planets (OPs) ($m$ 50% rocks, 50% water) upper curve, and rocky planets ($m$ 100% rocks) lower curve, as obtained from models of their internal structure by Sotin et al. (2007) when planets are located in the habitable zone of their star, or further. For a given mass, the mean densities and the radii of the planets are significantly different according to their nature. Planetary mass and radius are accessible to observational measurements by Doppler velocimetry and planetary transit photometry, respectively. These measurements have the potential capability of discriminating between the two types of planets. It should be noted that some of the Galilean satellites of Jupiter are examples of OPs, but for their lower masses.

### 1.2. On the verge of detecting ocean-planets?

Thanks to remarkable progress, radial velocity (RV) measurements are now unveiling a population of planets with a mass between 10 and 20 $M_{Earth}$ that are found at short orbital distances and as far as the habitable zone of their star. This is well illustrated by the triple system HD69830a,b,c (Lovis et al., 2006). Models reproducing this system invoke the formation of icy–rocky cores beyond the snow line, followed by inward migration (Alibert et al., 2006; Terquem and Papaloizou, 2006). Because of their higher mass, these *hot Neptunes* differ from OPs by their H–He envelope of several $M_{Earth}$. Cores above 10 $M_{Earth}$ are indeed expected to accrete a significant fraction of their mass as hydrogen-rich gas.

For two reasons, these discoveries are extremely promising for the search of OPs: First, the same mechanism that commonly produces *hot Neptunes* should also generate OPs for less massive migrating rocky–icy cores. Second, the increased accuracy in RV measurements should soon allow us to detect planets below 10 $M_{Earth}$, in the range of masses of OPs.

The CoRoT mission (http://CoRoT.oamp.fr; Rouan et al., 1999) was successfully launched by the end of 2006 and *Kepler* is scheduled for launch in 2008 (http://kepler.nasa.gov; Koch et al., 2006). Their transit search programs will determine the radii of discovered planets. The Doppler follow-up, whenever possible, will determine their masses. For each detected planet, these observations will determine a point and its box error in the ($M_{pl}$, $R_{pl}$) plane. Models of their internal structure predict $R_{pl}$ ($M_{pl}$) curves in that plane, which vary with the planetary composition (Fig. 1). Assuming that the models are correct, the relative positions of the error box and the curves will lead, or not, to the characterization of the nature of the planet. This approach was successfully applied to transiting *hot Jupiters*, among which we can distinguish a core-dominated planet (HD 149026b; Sato et al., 2005) from the other gas-dominated planets (for instance TrEs-1; Alonso et al., 2004). In return, the observations have put serious constraints on the models and the structure of hot Jupiters, or *Pegasides*. For instance, the large radius of HD209458b is not yet well explained.

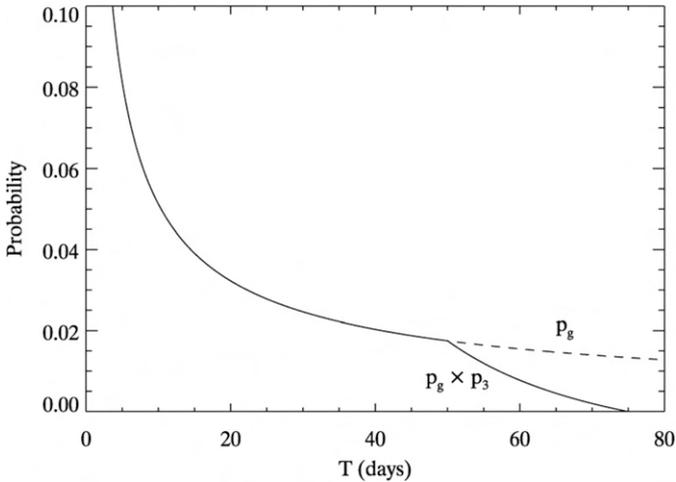

Fig. 2. Probability of observing an existing planet around a target star by CoRoT, as a function of its orbital period. Assuming a random orientation of planetary orbits, the geometrical probability that a transit is observed from the Earth is $p_g = R_{st}/a$, the ratio of the stellar radius to that of the planet distance to its star, in the limit $R_{pl} \ll R_{st}$. Using the Keplerian relation between $a$ and the planetary period $T_{orb}$ (for a given stellar mass, $T_{orb}^2/a^3 = $ const) and the time probability that $\geqslant 3$ events occur within a 150 day duration, $p_3$, one obtains the probability that a given planet around a GV star is detected. This is a rapidly decreasing function of the planetary period, which points out that planet detection is biased towards short period objects.

*1.3. Planet orbital periods accessible to CoRoT and Kepler*

CoRoT was initiated as a CNES "small mission." It is located in a low-altitude Earth orbit (~900 km). As it must not point too far from the anti-solar direction, it can observe a given stellar field continuously only during 5 months (150 days). It is considered that the detection of a planetary candidate by the transit method requires the observation of 3 transits or more. Thus, only planets with period $P < 75$ days can be detected (Fig. 2). For a solar-type star, this corresponds to a distance to its star, $a$, less than 0.35 AU (circular orbit) and a blackbody temperature (albedo = 0, no greenhouse effect) $T_{bb} > 460$ K.

Planets with $P < 75$ days can be habitable if their host-star is less massive than a K5 star ($M \sim 0.7\ M_{Sun}$). However, these low-mass stars represent only 1.5% of the stars accessible with CoRoT. With less than 200 of them in its field and a transit probability of ~1%, CoRoT is unlikely to detect the transit of a habitable planet.

Statistically, the probability that 3 transits or more of a given planet are detected within 150 days is a rapidly decreasing function of its orbital period (Fig. 2). If planets were uniformly distributed in distances around their stars, and if CoRoT had the sensitivity to detect all of them, the histogram of detections would be proportional to this probability, stressing the strong bias towards short periods. In addition, the larger the number of transits the larger the detection S/N for small-size planets. As a consequence, it is important to address the fate of OPs when they are close to their stars.

The situation for the *Kepler* mission is more favorable (http://kepler.nasa.gov; Koch et al., 2006). It is a larger mission that will operate on an Earth-trailing heliocentric orbit, continuously for 4 years. It can detect planets with period $\leqslant 1.33$ yr around solar-type stars, corresponding to semimajor axis $a \leqslant 1.21$ AU and $T_{bb} \geqslant 250$ K, which includes Earth-like planets, the main goal of the mission. However, even for *Kepler*, the detection of inner planets will be easier and more accurate.

According to Bordé et al. (2003), CoRoT could detect several tens of planets with $R_{pl} \sim 2\ R_{Earth}$ and $a \times (L/L_{Sun})^{1/2} < 0.35$ AU, where $L$ is the stellar luminosity, *if* each star has one planet within these ranges of size and location. This hypothesis is arbitrary, but it indicates that the actual number of detections could tell us the abundance of these planets if the mission capacities are as expected. For *Kepler*, the prospects are higher and start from planets with $a \leqslant 1.21$ AU for Sun-like stars.

*1.4. Outline of this study*

Due to the observation methods described above, exoplanets with the shortest periods are the easiest to characterize by their mass and radius. On the other hand, and as already discussed by Kuchner (2003), the strong stellar irradiation can affect the evolution of the water reservoir of close-in planets (by atmospheric escape) and of their radius. In summary, OPs that have kept enough water to be distinguished from silicate-dominated planets may be found only at orbital period that will not be available soon to accurate mass and radius measurements. This question provides the general outline of our paper.

In Section 2, we address the fate of water-rich planets that have migrated very close to their star, their swelling due to the strong external heating, and, especially, the loss of their water through thermal and non-thermal evaporation induced by stellar X and EUV radiation, wind and coronal mass ejections (CMEs).

In Section 3, we estimate the accuracy of the radius and mass determination, as a function of the orbital period and stellar properties, that can be obtained with CoRoT, *Kepler*, and RV follow up. Then, on the basis of this accuracy and of the theoretical $R_{pl}(M_{pl}, w)$ curves, we evaluate the capability to distinguish a water-rich from a silicate-rich population of planet.

In Section 4, we discuss what the realistic diversity of 1–10 $M_{Earth}$ short-period planets could be. In particular we address the possible degeneracy of ($R_{pl}$, $M_{pl}$) couples of values due to the existence of an accreted atmosphere of $H_2$–He, and to the existence of planets with a different bulk composition such as *carbon-planets* (Kuchner and Seager, 2005). Our conclusions are presented in Section 5.

## 2. Structure and fate of a short period ocean-planet

OPs are defined as planets that form farther away than the snow line and migrate inwards. When the orbital distance stabilizes within the habitable zone, and assuming that the planet has cooled to the thermal equilibrium with the stellar irradiation, a global and thick ocean would condense above an icy mantle (Léger et al., 2004). For shorter orbital distances no gas/liquid interface can be sustained. Assuming that water is the main constituent of the atmosphere, this happens when there is no more phase transition between the vapor phase and the liquid one, i.e. when the temperature distribution within the planet $T(P)$ runs above the critical point of water ($T_c = 647$ K, $P_c = 22$ MPa $\sim 220$ bars) as illustrated in Fig. 3a. When an ocean is present, the $T(P)$ curve crosses the gas–liquid transition line, which determines the ocean surface location (Fig. 3b). In the absence of liquid–gas interface, the whole fluid layer can be called *envelope* as in giant planets. The whole water envelope is supercritical. These hot planets could be named "planets with supercritical water envelope" or "sauna-planets." For the sake of simplicity, we shall keep the name of hot ocean-planets.

The possibility that transit missions detect and characterize some OPs depends on their existence/abundance and the instrument capabilities to detect small planets, say with $R_{pl} \sim 2 R_{Earth}$, at small distances from their stars, as discussed in Section 3. In the present section, we discuss whether the $R_{pl}(M_{pl})$ relation established by Sotin et al. (2007) (Fig. 1) in the HZ is modified by a swelling of the planet due to stronger irradiation (Section 2.1) and how close to their star OPs can be without losing most of their water (Section 2.2).

### 2.1. Estimated swelling of an ocean-planet close to its star

Sotin et al. (2007) and Valencia et al. (2006) modeled the internal structure of OPs and established the $R_{pl}(M_{pl})$ relation for OPs in the HZ. When these planets are closer to their stars, the higher stellar flux induces a higher temperature at the outer boundary, which may result in an enhanced radius. In order to test the sensitivity of the radius to this outer boundary temperature, Sotin et al. (2007) have calculated that, in the case of a 5 $M_{Earth}$ OP, an increase of 1000 K results in a 0.9% increase of the radius. However, this computation was done only for the condensed (solid and liquid) part of the planet. A more significant swelling is expected from the expansion of the hot water vapor atmosphere. Inside the HZ, the atmospheric water vapor content is a function of the orbital distance: around the present Sun, in the absence of other greenhouse gases and assuming a large ocean at the surface, the partial pressure of $H_2O$ would be 1 bar at 0.93 AU and would reach the critical pressure (220 bar) at 0.84 AU (Kasting, 1988). At closer distances, the outer part of the planet is a thick supercritical water envelope, or "steam-ocean," bounded by the high-pressure ice mantle, or by the silicate mantle if the temperature imposes that all the water is fluid. The depth of an hot "steam-ocean" would be larger than the depth of liquid–solid water mantle of the same mass, while the subcritical atmosphere above $P = 220$ bar expands quasi-linearly with its temperature and might thus result in a swelling.

Let us first consider the swelling due to the expansion of the opaque atmosphere. The pressure in the atmosphere depends

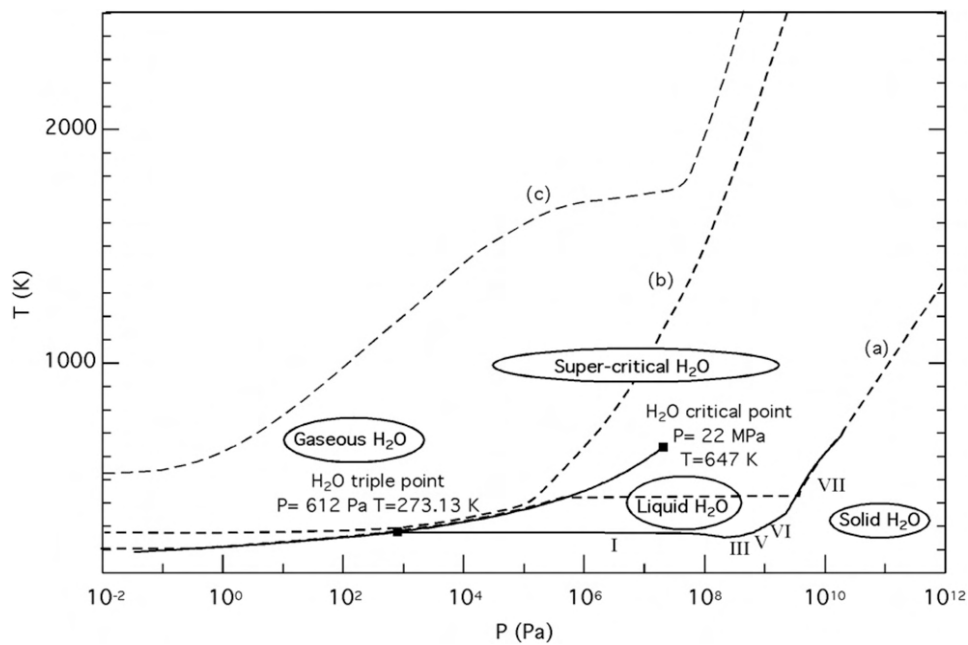

Fig. 3. Water phase diagram and *qualitative* $T(P)$ profiles for the outer layers of a water-rich planet. Planet (a) is in the habitable zone (HZ) of its star and has an ocean–atmosphere interface. Planet (b) is at an orbital distance slightly shorter than the inner edge of the HZ and planet (c) has a very hot equilibrium temperature (>500 K). Planets (b) and (c) have no ocean–atmosphere interface but a thick fluid envelope of supercritical $H_2O$.

upon the altitude, $z$, as

$$P = P_0 e^{-z/H}, \quad H = CkT/\mu g, \qquad (1)$$

where $H$ is the atmospheric height scale, $P_0$ the critical pressure at the lower boundary, $k$ the Boltzmann constant, $\mu$ the mean molecular weight, $g$ the local gravity, and $C$ is the compressibility factor, which depends on the pressure and temperature, $C = 1$ for a perfect gas and $C < 1$ for a dense gas.

The radius determined by a transit observation is the distance $z$, to the planetary center, of a light ray grazing the planetary atmosphere with an optical thickness of unity at the relevant wavelength. Wavelength-dependent radii of transiting Earth-size planets have been calculated by Ehrenreich et al. (2006). They present transmitted spectra of transiting ocean-planets with or without clouds. For a hot ocean-planet, water clouds are not expected because the gas is at a temperature $\geqslant 1000$ K (see below), which is above the critical point of water and prevents the condensation of liquid droplets. Assuming that there are no other aerosols or grains (Fe or silicates), in the upper atmosphere, Rayleigh scattering is expected to be the main contributor to the total extinction. An approximation of the column density $N$ across the atmospheric limb along the line of sight is $N = n(2\pi R_{\rm pl} H)^{1/2}$, where $n = P/(kT)$ is the volumic density (Fortney, 2005). The corresponding optical depth has been calculated by Ehrenreich et al. (2006) and can expressed as

$$\tau(\lambda) = \sqrt{2\pi} \frac{P(z)}{kT} R_{\rm pl}^{1/2} H^{1/2} \sigma_{\rm Rayl}(l), \qquad (2)$$

where $\sigma_{\rm Rayl}$ is the Rayleigh cross-section of the gas molecules (here water). Requiring that $\tau = 1$ at 0.6 μm, a central wavelength of the photometric measurements, implies a value for $P$ and then $z$ from Eq. (1). Normalizing to a reference case, using $\sigma_{\rm Rayl}$ (H$_2$O, 0.6 μm) $= 2.32 \times 10^{-27}$ cm$^2$, one reads:

$$P = 0.38 \bar{T}^{1/2} \bar{M}^{1/2} \bar{R}^{-3/2} \text{ bar}, \qquad (3)$$

with

$$\bar{T} = T/10^3 \text{ K}, \quad \bar{M} = M_{\rm pl}/6 M_{\rm Earth}, \quad \bar{R} = R_{\rm pl}/6 R_{\rm Earth}.$$

For the reference case, one finds $z - z_0 = 13$ km where $z$ and $z_0$ are the altitudes where $\tau = 1$ at 0.6 μm ($P = 0.38$ bar), and $P_0 = 1$ bar, respectively. The effective length $l_{\rm eff}$ of the light travel in the atmosphere is then 1000 km, corresponding to a sustaining angle $2\theta = 4.3°$. For comparison, for a transiting Earth, using $\sigma_{\rm Rayl}$ (air, 0.6 μm) $= 3.16 \times 10^{-27}$ cm$^2$ and neglecting the absorptions due to O$_3$ and the aerosols, one gets: $P = 0.24$ bar and $z - z_0 = 11.4$ km, $l_{\rm eff} = 640$ km, $2\theta = 5.8°$. These values have to be compared with the much higher attenuation of the Sun at sunset when the value of $P$ is 1 bar and only half of the grazing travel is performed by the light.

In the following estimates of the swelling of a hot OP, we consider that the planetary radius determined by a transit observation corresponds to $P \sim 0.4$ bar, and we determine the height of the atmospheric layer bounded by the 0.4 and 220 bar levels.

At the orbital distance of a hot Jupiter (0.05 AU), the temperature of the 0.4 bar level of an OP would be close to its equilibrium temperature, that is 1100 K. In a hot Jupiter of with a similar equilibrium temperature, the transition between the outer radiative layer and the convective region occurs in the H$_2$–He envelope at pressures and temperatures of the order of 500–1000 bar, 2000–3000 K (Barman et al., 2005). Although we would need a self-consistent model to compute the $T$–$P$ profile of a water envelope, as a function of the orbital distance and taking into account the precise opacities of H$_2$O, we can assume that the atmospheric temperature increases from about 1100 K in the upper and optically-thin layers, to typically 2000–2500 K at the transition between the radiative and convective region, that we arbitrarily fix at 200 bars here. By estimating the height of a 1000 and 2500 K isothermal atmosphere at hydrostatic equilibrium, we can bracket the height of the radiative layer that is opaque to the transit observer. For a 6 $M_{\rm Earth}$ planet and these two atmospheric temperatures, the 0.4 bar level is reached, respectively, at 260 and 700 km above the 220 bar level. These estimates take into account the non-ideal properties of the high pressure water vapor (NIST steam tables for the compressibility factor), but one can note that using the perfect gas approximation yields similar values (210 and 700 km).

The intrinsic swelling of the convective envelope underneath the radiative atmosphere is lower but it affects a large fraction of the radius. The problem to estimate the depth of this layer as a function of the temperature of the outer layers is double. First we are rapidly limited by the lack of available data for the equation of state of water at high temperature and pressure. Then, as the cooling of the planets is determined by the gradient of temperature in the outer layers, a high equilibrium temperature of the planet can lead to a significantly decreased heat flux and very hot interior. Addressing this issue would require a consistent formation model providing initial thermal conditions and an evolution code. Very preliminary estimate using extrapolations of available data show that the depth of the convective envelope of an hot 6 $M_{\rm Earth}$ can increase by $\sim$300–500 km.

Compare to the 2 $R_{\rm Earth}$ (12,800 km) radius it would have in the HZ, a hot 6 $M_{\rm Earth}$ OP (at 0.05 UA from its star) would exhibit a radius increased by 5–10%: +200–700 km for the atmosphere, +300–500 km for the fluid water envelope, +<100 km for the silicate + metal interior. For smaller planets and planets still closer to their stars, the swelling can be somewhat larger. For 1 $M_{\rm Earth}$, at 0.05 AU, the swelling should be >10%.

Rocky planets could also experience such a swelling of their atmosphere, in principle making them difficult to distinguish from OPs. However, at the orbital distances where the swelling could be significant, silicate-dominated planets would lose rapidly their volatile contents through thermal and non-thermal atmospheric loss processes, while OPs have an almost inexhaustible reservoir of water (see below).

Hence, when an OP is close to its star, it may undergo some swelling that would help to discriminate it from a rocky planet. However, as these estimates for the swelling are very preliminary and concern only very-short-period planets we will assume conservatively that, if it survives, an OP will have a radius similar to what it would have if it were located within the HZ.

During the planet formation process, some hydrogen may be accreted, and water is likely to be in contact with reducing Fe$^{2+}$

ions, possibly even metallic iron; hence, one can expect a chemical reaction with a hydrogen output. The presence of $H_2$ would imply a larger swelling of the planet. However, the atmospheric erosion processes that are considered in the next sections are much more efficient for hydrogen than for the other gases and will rapidly take hydrogen away from the atmosphere, especially at the orbital distances relevant for CoRoT. Thereafter, we consider that OP's atmospheres are mainly free of hydrogen, but the actual hydrogen content of OPs merits further investigations.

### 2.2. Evaporation of short period ocean-planets

Ocean-planets close to their stars may lose their water to space and become rocky planets. The aim of the present section is to estimate how close to their stars we can expect OPs to keep their water and possibly be identified by CoRoT and *Kepler*. The precise modeling of the erosion of an OP is complex, and the corresponding work is still to be done. We can, however, estimate the upper limit on atmospheric losses based on maximum energy deposition in the upper atmosphere. Using energy-limited escape, which is likely to be significantly larger than actual loss rates, we can determine the minimum period (or orbital distance) at which most of the water reservoir should survive during a given time, for instance 5 Gyrs. Determining this minimum orbital distance is important in the context of CoRot searches that will focus on close-in planets. Note, however, that planets found at shorter periods could still have kept most of their water, depending on whether actual losses are limited by mechanisms other than energy deposition.

The main processes for atmospheric escape that can lead eventually to the exhaust of the planetary water reservoir are: (1) thermal escape driven by exosphere heating by X-rays and Extreme UV (or XUV: 0.1–100 nm), (2) non-thermal escape driven by the action of particles from the star.

#### 2.2.1. Water erosion by thermal escape

The heating of the upper atmosphere by XUV photons can result in the escape of gases to space. Escape affects mainly light species like H and He, but heavier species like O can also be carried away by the hydrodynamic flow when it is high enough (Chasseféire, 1996): such an effect has been observed in the case of the hot Jupiter HD209458b (at 0.05 AU from its star) for which not only H, but also C and O, have been detected in the escaping upper atmosphere (Vidal-Madjar et al., 2003, 2004). An upper limit to the mass loss is given by the energy-limited escape rate (Lammer et al., 2003), which would be reached if all the energy absorbed at $\lambda < 100$ nm would be converted into gravitational energy, as expressed in the following equation

$$\varepsilon \frac{F_{XUV} \pi R_{pl}^2}{(a/1 \text{ AU})^2} = \frac{GM_{pl}\dot{m}_{H_2O}}{R_{pl}}. \quad (4)$$

The left term is the fraction of the XUV flux that is intercepted by the planet and available for driving the thermal escape, and the right term is the variation of gravitational energy induced by the mass-loss of water. $F_{XUV}$ is the XUV energy flux [Watt m$^{-2}$] received at 1 AU, $M$, $R$ and $a$ are the mass, radius and orbital distance of the planet, respectively; the factor $\varepsilon$ corresponds to the efficiency of the conversion of incident XUV energy into effective escape of gas, and contains all the physical complexity of the escape process. The mass of water lost over a time $t$ is

$$m_{H_2O} = \frac{\varepsilon \pi R_{pl}^3}{GM_{pl}(a/1 \text{ AU})^2} \int_0^t F_{XUV}(t')\,dt'. \quad (5)$$

We can consider $\varepsilon$ as the product of a heating efficiency $\varepsilon_h$, which gives the fraction of the incident energy that is not re-radiated to space, and an escape efficiency $\varepsilon_{esc}$, which is the fraction of the deposited energy that is eventually lost through escaping gas. The heating efficiency, $\varepsilon_h$, is typically lower than 0.2, especially at high XUV irradiation. Yelle (2004) used a detailed photochemical model and found a value of ∼0.1 for HD209458b. Here, we use a value of 0.2 in order to obtain an upper limit on the loss rate.

In the preceding relations, $R_{pl}$ is sometimes replaced by the radius at which the XUV radiation is absorbed $R_{XUV}$ (e.g. Lammer et al., 2003). This comes from Watson et al. (1981) hydrodynamic modeling of the escape in which all the incoming XUV radiation is assumed to be absorbed in an infinitely thin layer at $R_{XUV}$. In this approach, $R_{XUV}$ can be as large as several times $R_{pl}$, for terrestrial planets. This expansion of the upper layer can imply extremely high escape rates. Recent hydrodynamic models (steady state: Tian et al., 2005; time dependent: Penz et al., 2006) including a realistic deposition of the XUV along the escaping flow, show that the thin absorbing layer approach overestimates the escape rate by orders of magnitude, especially for high incoming XUV flux. One of the reasons for this overestimation is the XUV self-shielding: some XUV photons that are absorbed within the escaping outflow contribute to heating the outflow but not to enhancing the loss rate. When the XUV deposition is just high enough to drive hydrodynamic escape, the escaping outflow is not dense enough to absorb a significant fraction of the incoming XUV energy, which is mostly deposited deeper where part of it is eventually converted into gravitational energy. In this phase, the escape rate increases nearly linearly with the XUV flux, but with increasing irradiation, escape reaches a point at which the column density of escaping atoms can no longer be considered transparent to XUV: the outflow becomes hotter but the efficiency $\varepsilon_{esc}$ decreases.

The escape efficiency was estimated by Penz et al. (2006) by modeling the hydrodynamic flow. For HD209458b they found an escape efficiency of about 0.5 at 0.045 AU from the present Sun (conditions assumed for HD209458b) that decreases rapidly for increasing XUV fluxes ($\varepsilon_{esc} = 0.15$ if the XUV flux is multiplied by 2). Therefore, the total efficiency $\varepsilon = \varepsilon_h \varepsilon_{esc}$ varies from 0.02 to 0.1 in the range of orbital distances that are observable by CoRoT.

Fig. 4 gives the time required to lose the whole water reservoir of an OP through this energy-limited mass-loss as a function of the orbital distance $a$ (solid lines). This estimate takes into account the evolution of the XUV stellar luminosity ac-cording to Ribas et al. (2005) and the variation of $\varepsilon$ with the XUV flux from Penz et al. (2006). For OPs with a mass below 6 Earth masses, the loss is limited by the XUV energy deposition. For a 5 Earth-mass OP, Fig. 4 shows that the thermal loss of the water reservoir (H and O) would take more than 5 Gyrs at $a \geqslant 0.04$ AU.

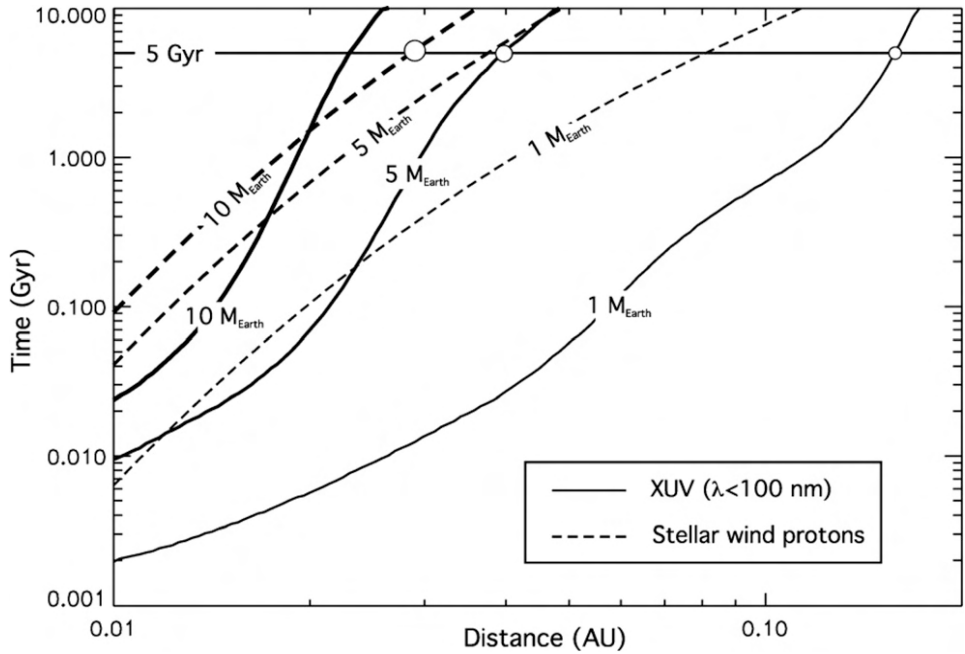

Fig. 4. Minimum lifetime of the water content of a 10, 5 and 1 Earth-mass OP as a function of its orbital distance. Solid lines give the lifetime of the water reservoir eroded by thermal escape at its energy-limited rate (determined by the deposition of stellar XUV radiation). Dashed lines give the energy-limited lifetime for non-thermal escape induced by the stellar wind (determined here by the deposition of protons' kinetic energy). See Section 2.2 for the assumptions on the efficiency of the thermal escape, on stellar XUV and wind properties. Around a 5 Gyr old star, OPs would be weakly affected at orbital distances $\geqslant 0.03$ AU (10 $M_{Earth}$), 0.04 AU (5 $M_{Earth}$) and 0.15 AU (1 $M_{Earth}$).

*2.2.2. Water erosion by the stellar wind and coronal mass ejections*

Now, the question is whether the erosion by the stellar wind can significantly change the lower limit of the water reservoir lifetime. This process is efficient when the upper atmosphere of a planet is not protected by an intrinsic magnetic field, which is strong enough to deflect the stellar plasma flow at large distances above the exobase.

OPs close by their stars are expected to have their spin rotation tidally locked on their orbital rotation. From simple theoretical models, several analytical scaling laws were derived which allow one to estimate the planetary magnetic dipole moment from the planetary characteristics (e.g. density, rotation rate, size of the dynamo region). These models are summarized and compared in Grießmeier et al. (2004, 2005). According to these models, the expected magnetic moments of slowly rotating exoplanets, e.g. tidally locked planets, are smaller than that of fast rotating planet like the Earth. By applying these scaling relations to an OP of 6.0 Earth masses and 2.0 Earth radii (Léger et al., 2004), which is tidally locked at 0.2 AU around a star with 0.2 solar masses, we obtain a maximum intrinsic magnetic moment of 0.7 times that of the present Earth. Because of the relatively weak magnetic moment (considering the large planetary radius), the planetary magnetosphere may not be efficient at protecting the planetary atmosphere.

In such a Venus-like interaction (e.g. Terada et al., 2002) the planetary neutral gas in the upper atmosphere can be ionized by electron impact, X-rays and Extreme UV radiations. After charge exchange during collisions with the incident stellar plasma, it is picked up by this stellar plasma flow and escapes from the planet.

In the two next sections we estimate the non-thermal atmospheric loss due to the interaction with the mean stellar wind and with the sporadic coronal mass ejections (CMEs), integrated over the whole history of the system.

*Stellar wind* As an upper limit to the erosion due to the stellar wind, we calculated the flux of protons intercepted by the planet *in the absence of magnetic deflection* and the corresponding deposition of kinetic energy into the upper atmosphere. By analogy with the energy-limited thermal loss rate described in Section 2.2.1, we can infer an energy-limited escape induced by the stellar wind protons. The main difficulty with this approach is the poorly known evolution of the stellar wind. Using indirect observations, Wood et al. (2002, 2005) showed that the mass loss of Sun-like stars is correlated with their X-ray luminosity for ages above 0.7 Gyr. Younger stars do not seem to exhibit a higher mass loss. In order to make an upper estimate of the possible losses, we used constant wind properties corresponding the highest mass loss measured by Wood et al. (2005). This allows us to overestimate the energy deposition. An improvement of this assumption would require more data on stellar winds, especially on those of young stars. Dashed lines on Fig. 4 show the result we find for the minimum lifetime of the water reservoir. At a distance $a \geqslant 0.04$ AU, a 5 Earth-mass OP would keep its water for more than 5 Gyrs if stellar wind erosion is considered alone.

This is an upper limit for the non-thermal loss through mean-stellar-wind protons, but the effect of coronal mass ejections (CMEs) from the star has also to be addressed.

*Coronal mass ejections*   By using the parameters estimated by Khodachenko et al. (2007) one obtains at 0.05 AU, an average CME plasma velocity of $\sim 500$ km s$^{-1}$ and plasma (proton) densities $\leqslant 5 \times 10^4$ cm$^{-3}$.

Lammer et al. (2006) studied maximum rates of O$^+$ ion pick up by CME for magnetized and non-magnetized Earth-like exoplanets. They calculated the plasma flow around the magnetopause/ionopause for XUV flux values 10–100 times higher than that of the present Sun, as expected for a young star (Ribas et al., 2005). The total ion pick up loss rate was obtained by calculating the O$^+$ production rate caused by charge exchange, electron impact and photons along the streamlines in the plasma flow around the planetary obstacle.

They found that for strong CMEs and for the estimated minimum planetary magnetic moment, the situation would result in a Venus-like plasma interaction (e.g. Terada et al., 2002). Putting in the typical numbers for the CME proton density and the average wind speed mentioned above, we find that, during the critical first 0.5 Gyr for an OP at 0.05 AU, the water losses could be up to $\leqslant 200$ Earth oceans (by mass), which is significantly less than the losses by stellar winds and can be neglected.

One should note that these estimates are also *upper limits* because the more massive OP considered in our study has a higher gravitational potential than the Earth-mass planets studied by the preceding authors, which will hold back the thermosphere–exosphere environment expansion, resulting in lower loss rates. Furthermore, the evaporating hydrogen will form a corona around the planet that also decreases the non-thermal loss of oxygen and other heavy species by protecting them.

*Those upper estimates* of non-thermal loss processes from weakly magnetized OPs *do not come close to exhausting the important water reservoir of an OP that is located farther than 0.05 AU*. For instance, 200 Earth oceans ($200 \times 1.45 \times 10^{21}$ kg) represent only 1.5% of the mass of a 6 Earth-mass OP.

### 2.2.3. How close to its star can we expect to find an ocean-planet?

According to Fig. 4, non-thermal escape induced by the stellar wind seems to dominate the atmospheric loss for $M_{pl} \leqslant 5\,M_{Earth}$. For larger masses, the survival of the water reservoir is longer than 5 Gyrs. In the absence of more detailed models for the stellar wind-induced losses, our assumptions (100% efficiency, very intense stellar wind, no intrinsic magnetic field) certainly overestimate significantly the solar wind-induced loss (Section 2.2.2). With these over maximizing assumptions, finding non-thermal losses much higher than the thermal escape would have prevented us from providing a firm conclusion on the survival of OPs. Fortunately, we find an upper limit on the solar wind induced loss that is lower, or only slightly above, the thermal loss. We can thus derive an estimate for the closest distance at which OPs will not be affected by water loss to space: as a typical result, *a 6 $M_{Earth}$ OP will keep most of its water when it is farther than 0.04 AU from its star*.

This conclusion is very similar to the results obtained by Kuchner (2003), who already studied the survival of the volatile content of short period OPs: on Fig. 4 of his paper one can see that the lifetime of 5 $M_{Earth}$ *volatile-rich planet at 0.04 AU is 5 Gyr*. But Kuchner also pointed out 2 potential problems affecting his calculations at the shortest orbital periods.

The first one is related to the $R_{XUV}/R_{pl}$ ratio (discussed in Section 2.2.1) that strongly overestimates the loss rate at high XUV irradiation in Watson's numerical approach of the hydrodynamic escape. To estimate the efficiency the loss at high XUV irradiation, we used recent hydrodynamic calculations (Penz et al., 2006) and found losses much below what is given by Watson's scheme. It should, however, be noted that the value of this efficiency has a strong impact on the determination of the minimum distance, and that hydrodynamic modeling of atmospheric escape is a field in progress. Different estimates of this efficiency might be obtained in the future.

A second point raised by Kuchner is that the atmospheric loss can induce an expansion of the whole gaseous envelope. Because of positive feedback on the loss rate, this effect could lead to a runaway escape and to the loss of the whole volatile reservoir within a much shorter timescale than what is found here by assuming no coupling between the hydrostatic structure and the mass loss. This runaway loss was studied for gas giants by Baraffe et al. (2004) and was shown to occur when the timescale of the loss $m/\dot{m}$ becomes shorter than the thermal (Kelvin–Helmholtz) timescale required for thermal readjustment of the envelope to a modification of its mass. This condition can be written:

$$\frac{m}{\dot{m}} < \frac{Gm^2}{R_{pl}L_{pl}}. \tag{6}$$

Here, $m$ is the mass of the water envelope and $L_{pl}$ is the planet luminosity, which can be written as $4\pi R_{pl}^2 \sigma T_{eff}$ by assuming a mean brightness temperature $T_{eff}$, or $2\pi R_{pl}^2 \sigma T_{eff}$ if there is no redistribution of the incoming energy between the day and night hemispheres. The runaway condition can be written

$$\frac{2\sigma T_{eff}^4}{\varepsilon F_{XUV}} < 1 \quad \text{or} \quad T_{eff} < \left(\frac{\varepsilon F_{XUV}}{2\sigma}\right)^{1/4}, \tag{7}$$

where $\varepsilon$ is the conversion efficiency. For the closest orbital distances we considered, this condition for the runaway implies $T_{eff} < 240$ K. At these short orbital distances, realistic effective temperatures cannot drop below 500 K, and we can assume that thermal readjustment are instantaneous compared to the loss timescale, which is compatible with our quasi-static/hydrostatic approach. We conclude that the runaway loss of water is not expected. However, we stress again the central influence of the efficiency $\varepsilon$. Higher efficiencies than the one obtained by Penz et al. (2006) could trigger a runaway loss and the loss of the whole water reservoir within short timescales, as described in Baraffe et al. (2004).

## 3. Could $R_{pl}$ and $M_{pl}$ measurements be accurate enough to discriminate ocean-planets from rocky planets?

In order to know whether an exoplanet is an ocean-planet or a rocky one, its mass and radius must be measured with high enough accuracy so that the position of the error box in the ($M_{pl}$, $R_{pl}$) plane allows a distinction between the predictions for those two types of planets, assuming that the model is adequate (Fig. 1). For small masses and radii ($M_{pl} < 10\,M_{Earth}$, $R_{pl} \leqslant 2\,R_{Earth}$), this is a not a trivial task. Hereafter, we estimate the different uncertainties that define the error boxes.

### 3.1. What are the different sources of uncertainty?

Around a given ($M_{pl}$, $R_{pl}$) point, the physical quantity that the planetary models predict to be different is the mean density. This density, $\langle\rho\rangle = (3/(4\pi))M_{pl}R_{pl}^{-3}$ can be obtained by proper observations. The planetary mass results from Doppler velocimetry. The planetary radius is obtained from the target star radius estimate and the relative stellar flux drop during the transit, i.e. $R_{pl} = R_{st}(\Delta F/F)^{1/2}$, if the stellar limb darkening can be neglected.

The uncertainties on the mass, the stellar radius estimate and flux measurements being independent, their variances add in quadrature. We estimate the uncertainty on the 1/3 power of the density, because this quantity has a relative uncertainty that is smaller than that on the density, which is more appropriate for linear expansions (Protanov, 2002):

$$\left(\frac{\sigma_{\rho^{1/3}}}{\rho^{1/3}}\right)^2 = \left(\frac{\sigma_{M_{pl}}}{3M_{pl}}\right)^2 + \left(\frac{\sigma_{R_{st}}}{R_{st}}\right)^2 + \left(\frac{\sigma_{\Delta F/F}}{2\Delta F/F}\right)^2. \quad (8)$$

Considering the case $M_{pl} = 6\,M_{Earth}$ that corresponds to a big telluric planet but that is still far from the limit where the hydrogen gas is likely to have been accreted, the internal structure modeling by Léger et al. (2004) yields radii of $R_1 = 2.0\,R_{Earth}$ and $R_2 = 1.63\,R_{Earth}$ for ocean-planets and rocky planets, respectively. The relative difference for the corresponding $\rho^{1/3}$ is:

$$\frac{\Delta\rho^{1/3}}{\rho^{1/3}} = \frac{\Delta R_{pl}}{R_{pl}} = \frac{2.0 - 1.63}{1.8} = 20.6\%. \quad (9)$$

Assuming a Gaussian distribution of errors, a $2\sigma$ characterization of $\Delta\rho^{1/3}/\rho^{1/3}$ would result in a 95% confidence discrimination between the two kinds of planets. Therefore, we consider that the standard deviation should satisfy:

$$\frac{\sigma_{\rho^{1/3}}}{\rho^{1/3}} \leqslant 10.3\%. \quad (10)$$

Next, the different contributions in relation (8) are estimated.

### 3.2. Uncertainty on the planetary mass

The semi-amplitude of the radial velocity wobble due to the presence of a planet around a solar type star is

$$K = 0.09\left(\frac{M_{pl}}{M_{Earth}}\right)\left(\frac{M_{st}}{M_{Sun}}\right)^{-1/2}\left(\frac{a}{1\,\text{AU}}\right)^{-1/2}\,\text{m s}^{-1}. \quad (11)$$

The corresponding uncertainty in the mass is

$$\frac{\sigma_{M_{pl}}}{M_{pl}} = \left[\left(\frac{\sigma_K}{K}\right)^2 + \left(\frac{\sigma_{M_{st}}}{2M_{st}}\right)^2\right]^{1/2}. \quad (12)$$

The uncertainty in the stellar masses of the targets stars that are considered ($m_v < 13$) can be estimated from that in the stellar radius, $\sigma_{R_{st}}/R_{st} \sim 5\%$ (Section 3.3), and $M(R)$ empirical relation (Cox, 2000): $\log(R) = 0.92\log(M) + $ const. The result is an ambiguity of $\sim 6\%$. This is a conservative value because, in practice, the mass is deduced from observable quantities using atmospheric models, whereas to obtain the radius an additional operation is needed that implies stellar evolution modeling. This result is in agreement with Charbonneau et al. (2006), who estimate that the uncertainty on the stellar mass can be as high as 5%. Anyhow, this term has a negligible contribution to the planetary mass uncertainty because it leads to $\sigma_{M_{st}}/2M_{st} \leqslant 3\%$, whereas $\sigma_K/K \sim 20\%$ (see below), and the hierarchy $(\sigma_{M_{st}}/2M_{st})^2 \ll (\sigma_K/K)^2$ always stands.

One of the very best Doppler instruments presently available is the HARPS spectrometer at the ESO-La Silla 3.6 m telescope. For faint stars ($m_v > 10$), its uncertainty is close to the fundamental shot noise limit. It will be the most accurate instrument for the follow-up of planetary transit candidates detected by CoRoT. In that program, the Doppler measurements will be performed in favorable conditions because the planetary ephemeris will be determined beforehand by transit photometry (transit period and epochs). The only unknowns will be the semi-amplitude of the sinusoidal radial velocity curve $K$ and the center-of-mass velocity $v_0$, if a circular orbit of the planet is assumed. The shot noise uncertainty of HARPS depends not only upon the stellar magnitude and the integration time, but also upon the spectral type and the rotational broadening $v\sin i$ of the observed star (Bouchy et al., 2001). For the most appropriate spectral types (from G5 to K5) and stars with low rotation ($v\sin i < 2$ km/s), the expected shot noise uncertainty on individual radial velocity measurements, $\sigma_V$, is given by Mayor et al. (2003) and Pepe et al. (2005):

$m_V = 14 \rightarrow \sigma_V = 6$ m/s (1 h),

$m_V = 12 \rightarrow \sigma_V = 2$ m/s (1 h).

For such stars, the uncertainties due to shot noise are larger than instrumental systematic errors (estimated to 0.8 m/s on HARPS). We suppose here non-active stars with surface velocities at the level of 2 m/s rms, which is the case of the most stable solar-like stars monitored with HARPS.

From $\sigma_V$ and the number of observations, the error $\sigma_K$ on $K$ can be computed as follows. In case of a circular orbit, the radial velocity reads

$$v(t) = v_0 + K\cos\left(\frac{2\pi t}{T_{orb}} + \varphi\right), \quad (13)$$

where the orbital period $T_{orb}$ and the phase $\varphi$ are known, thanks to transit photometry. Therefore, only two parameters, the systemic velocity $v_0$ and the semi-amplitude $K$, should be determined from measured velocities. Least-squares fitting would be the standard procedure for that purpose.

Following the proof in Section 15.2 of Press et al. (2002), and assuming that all measures have the same error $\sigma_v$, one can show that the variance of $K$ is

$$\sigma_K^2 = \frac{\sigma_v^3}{\sum_{i=1}^{N} \cos^2\left(\frac{2\pi t_i}{T_{\text{orb}}} + \varphi\right) - \frac{1}{N}\left(\sum_{i=1}^{N} \cos\left(\frac{2\pi t_i}{T_{\text{orb}}} + \varphi\right)\right)^2}. \quad (14)$$

If the measurements are uniformly distributed along the orbital period, the first sum can be approximated by $N$ times the integral of a squared cosine over its period, i.e., $N/2$, and the second sum can be approximated by the mean value of a cosine over its period, i.e., zero. Therefore,

$$\sigma_K^2 \approx \frac{\sigma_v^2}{N/2}. \quad (15)$$

However, if we take full advantage of the knowledge of the phase $\phi$ and perform the measurements in equal numbers at the maxima and minima of the velocity, then the first sum is equal to $N$ (because the squared cosine is always equal to 1), while the second sum is still zero. Hence

$$\sigma_K^2 \approx \frac{\sigma_v^2}{N}. \quad (16)$$

In practice, it might be difficult to take measurements only at the extrema of the velocity curve. A real case is well illustrated by Fig. 2 in O'Donovan et al. (2006), who performed their measurements close to the extrema. A possible value is

$$\sigma_K^2 \approx \frac{\sigma_v^2}{N/1.2}. \quad (17)$$

We assume that for an important project, a maximum integration time can be about 40 h, corresponding to a cumulative time of about 5 nights. In order to explicitly estimate the dependence of $\sigma_V$ upon the different parameters ($m_v$, $t$, $\Phi_{\text{tel}}$), we compute $(\sigma_V)_0$ for a given set of parameters and for a reference case, and we express $\sigma_V$ as a function of it and the parameters. In the shot noise-limited regime, the uncertainty on the radial velocity reads

$$\frac{\sigma_v}{(\sigma_v)_0} = 10^{0.2(m_v - m_{v_0})} \left(\frac{t}{t_0}\right)^{-1/2} \left(\frac{\Phi_{\text{tel}}}{\Phi_0}\right)^{-1}, \quad (18)$$

where $\Phi_{\text{tel}}$ is the telescope diameter, $t$ the cumulative integration time, and the index "0" corresponds to the reference case.

The reference case that we consider is: $M_{\text{pl},0} = 6~M_{\text{Earth}}$, $a_0 = 0.10$ AU, $M_{\text{st}} = M_{\text{Sun}}$. Relation (11) gives $K_0 = 1.7$ m/s. An $m_v = 12$ star observed with HARPS at the 3.6 m ESO telescope during $t_0 = 40$ h, leads to $(\sigma_K)_0 = 0.35$ m/s (Eq. (17)) and $(1/3)(\sigma_K/K)_0 = 0.068$. Using relations (18), the uncertainty on the planetary mass (relation (12)) reads

$$\frac{1}{3}\frac{\sigma_{M_{\text{pl}}}}{M_{\text{pl}}} = 6.9\% A \left(\frac{a}{0.10~\text{AU}}\right)^{1/2} 10^{0.2(m-12)}, \quad (19)$$

with

$$A = \left(\frac{M_{\text{pl}}}{6~M_{\text{Earth}}}\right)^{-1} \left(\frac{M_{\text{st}}}{M_{\text{Sun}}}\right)^{1/2} \left(\frac{t}{40~\text{h}}\right)^{-1/2} \left(\frac{\Phi_{\text{tel}}}{3.6~\text{m}}\right)^{-1}.$$

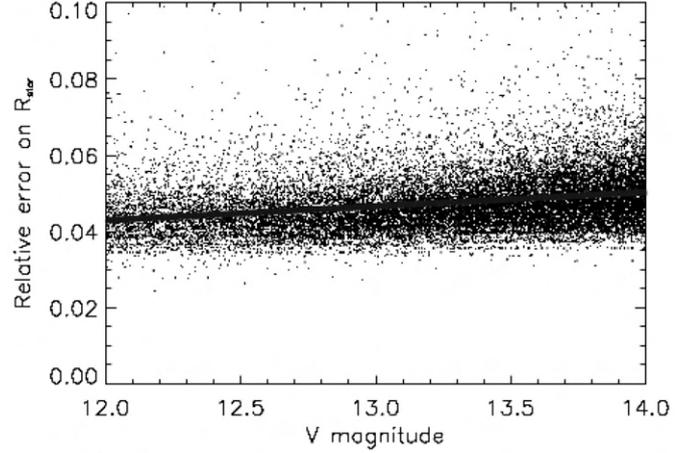

Fig. 5. Relative error on the stellar radius plotted against the magnitude of the CoRoT targets. The dots show the error estimated from the V and K actual magnitudes from relation (22) in Kervella et al. (2004), when the uncertainty on the reddening, or distance, is added. The full line shows the fit to the data.

### 3.3. Uncertainty on the stellar radius

The stellar radius of CoRoT targets will be estimated by in-depth follow-up observations of the parent star, including high-resolution high-SNR spectroscopy. A first idea is, however, obtained from calibrated relationships of color indices with stellar angular diameters (Kervella et al., 2004). Apparent magnitudes in the V and K filters are measured with a precision of typically 0.03 magnitudes in the CoRoT/exoplanet stellar catalogue. A preliminary spectral classification is then derived from broadband spectrophotometry in the visible and near-infrared, with an estimation of the reddening (Moutou et al., in preparation). High-resolution spectroscopy will then allow a more accurate estimation of the effective temperature, with error bars of typically 50 to 100 K (2–3% of the temperature). This will further constrain the reddening, at the level of typically an additional 0.05 magnitude uncertainty, which includes the error on distance estimation. The final error on the linear radius of the CoRoT targets is then obtained by combining an error on the angular diameter (deriving Eq. (22) in Kervella et al., 2004) with an error on the distance (from spectral type, reddening, absolute luminosity). Fig. 5 shows the calculated final error on the stellar radius for CoRoT targets in the magnitude range 12–14. Note that other methods of radius estimation give results that are consistent within 5% (Di Benedetto, 2005).

For comparison, the stellar radius of stars in known transiting systems is estimated with errors of typically 5–7% for OGLE systems (e.g. Santos et al., 2006). The expectations for CoRoT targets is of the order of what is achieved on OGLE targets of similar magnitudes, as a thorough follow-up observation program is foreseen to constrain at best the stellar parameters.

We conclude that the standard deviation on $R_{\text{st}}$ is estimated as

$$\sigma_{R_{\text{st}}}/R_{\text{st}} \sim 5\% \quad (20)$$

for stars brighter than $m_v \sim 13$, which implies a very careful photometric and spectroscopic analysis of the target, as inter-stellar extinction will be present in the line of sight of CoRoT targets.

### 3.4. Uncertainty on the amplitude of the transit

The amplitude of the relative flux variation during a transit, $\Delta F/F$, is $(R_{pl}/R_{st})^2$, or:

$$\frac{\Delta F}{F} = 0.84 \times 10^{-4} \left(\frac{R_{pl}}{R_{Earth}}\right)^2 \left(\frac{R_{star}}{R_{Sun}}\right)^{-2}. \quad (21)$$

For CoRoT, the uncertainty on the transit depth measurement is expected to be 1.2 times the shot noise for stars with $m_v =$ 11–14 (Fridlund et al., 2006). The shot noise depends on the instrument, the stellar magnitude, the number $k$ of transits and their individual duration $tr$. For a total duration observation, $t_{run}$, an orbital period $T_{orb}$, and an impact parameter (the ratio of the distance of the projected planet trajectory to the stellar center and the stellar radius) of 0.5, the cumulative transiting time, $t_{tot}$, is

$$t_{tot} = ktr, \quad (22)$$

with

$tr = 3.56(a/0.10 \text{ AU})^{1/2}(R_{st}/R_{Sun})(M_{st}/M_{Sun})^{-1/2}$ h,
$T_{orb} = 11.5(a/0.10 \text{ AU})^{3/2}(M_{st}/M_{Sun})^{-1/2}$ days,
$k = t_{run}/T_{orb}$.

$\Delta F$ is obtained by measuring the depth of the transit in the stellar light curve, $\Delta F = F_{tr} - F_{bg}$. The background term, $F_{bg}$, can be accurately estimated by averaging over long durations; therefore, we consider that the shot noise on $\Delta F$ is mainly that on $F_{tr}$. It is determined by calculating the number of photoelectrons at the flux level $F_{tr} \sim F$ (not at the level $\Delta F$) during the total transiting time. The uncertainty on $\Delta F$, in number of photoelectrons, is

$$\sigma_{\Delta F} = 1.2(Ft_{tot})^{1/2}. \quad (23)$$

The relative uncertainty on $F$ being much weaker than that on $\Delta F$, after some algebra, one reads

$$\frac{\sigma_{\Delta F/F}}{\Delta F/F} \approx \frac{1.2}{(\Delta F/F)(Ft_{tot})^{1/2}}. \quad (24)$$

CoRoT has a 0.27 m entrance pupil and observes during 150-day runs. It produces a $2.6 \times 10^4$ ph-el s$^{-1}$ flux for a 12th magnitude star (http://CoRoT.oamp.fr). Considering as a reference case, the transit of a $R_{pl} = 2 R_{Earth}$ planet in a circular orbit around a solar type star ($\Delta F/F = 3.36 \times 10^{-4}$), at distance 0.10 AU observed with CoRoT, the uncertainty is

$$\left(\frac{1}{2}\frac{\sigma_{\Delta F/F}}{\Delta F/F}\right)_0 = 2.7\%. \quad (25)$$

*Kepler* has a 0.95 m entrance pupil and observes during 4 years. For a 6.5 h duration transit and a 12th magnitude star, the differential photometry accuracy is estimated at 20 ppm, and is mainly shot noise limited for fainter stars (D. Koch, personal communication). For our planet-star reference case, for the whole mission duration, one finds

$$\left(\frac{1}{2}\frac{\sigma_{\Delta F/F}}{\Delta F/F}\right)_{Kepler} = 0.25\%. \quad (26)$$

In general, for a mainly shot noise limited observation, the uncertainty is

$$\frac{1}{2}\frac{\sigma_{\Delta F/F}}{\Delta F/F} = 2.7\% B \left(\frac{a}{0.10 \text{ AU}}\right)^{1/2} 10^{0.2(m_v-12)} \quad (27)$$

with

$$B = \left(\frac{R_{pl}}{2 R_{Earth}}\right)^{-2} \left(\frac{R_{st}}{R_{Sun}}\right)^{3/2} \left(\frac{t_{run}}{150 \text{ days}}\right)^{-1/2} \left(\frac{\Phi_{tel}}{0.27 \text{ m}}\right)^{-1}.$$

It is remarkable that the dependences of both uncertainties on the planetary mass (Eq. (19)) and transit amplitude (Eq. (27)) upon the orbital radius, $a$, and stellar magnitude, $m_v$, are the same.

### 3.5. Capabilities to identify the nature of the telluric-like planets

The different terms in Eq. (8) are now estimated. Using (19), (20) and (27), the condition for discrimination (Eq. (10)) reads

$$(0.05)^2 + [(0.069A)^2 + (0.027B)^2]\left(\frac{a}{0.10 \text{ AU}}\right) 10^{0.4(m_v-12)}$$
$$\leqslant (0.103)^2. \quad (28)$$

Considering the case of a planet with $M = 6\ M_{Earth}$, $R = 2\ R_{Earth}$ in front of a Sun-like star, observed with:

- CoRoT during 150 days and with 40 h of HARPS follow-up on a 3.6 m telescope, one reads $A = B = 1$. In the square bracket of Eq. (28), the term containing $A$ (the uncertainty on the mass) dominates that with $B$ by a factor 5. Although we have already considered a very long integration time for the Doppler measurement, the bottleneck remains the uncertainty on the planetary mass, even when using the small telescope of CoRoT;
- *Kepler* during 4 yrs and a 40 h follow-up with an HARPS-like spectrometer on a 3.6 m (8 m) telescope, one reads $A = 1$ (0.45), $B = 0.09$. Now, in Eq. (28), *the uncertainty on the mass fully dominates that on the radius* by a factor $\sim$800 (160). The way to improve the discrimination between the two types of planets is then *to improve the accuracy of the mass by using larger telescopes*.

This is under the assumption that, for faint stars, the uncertainties on $M_{pl}$ and $\Delta F/F$ are dominated by shot noise, rather than by the astronomical stellar noise or the instrument capabilities. A realistic detection exercise, including stellar noises in CoRoT simulated light curves, has shown that stellar micro-variability limits the detection only in extreme cases (Moutou et al., 2005); however, the impact of stellar micro-variability on transit parameter estimation is still under study.

Fig. 6 shows the regions in the $(m_v, a)$ plane where the discrimination between an ocean-planet and a rocky planet with 6

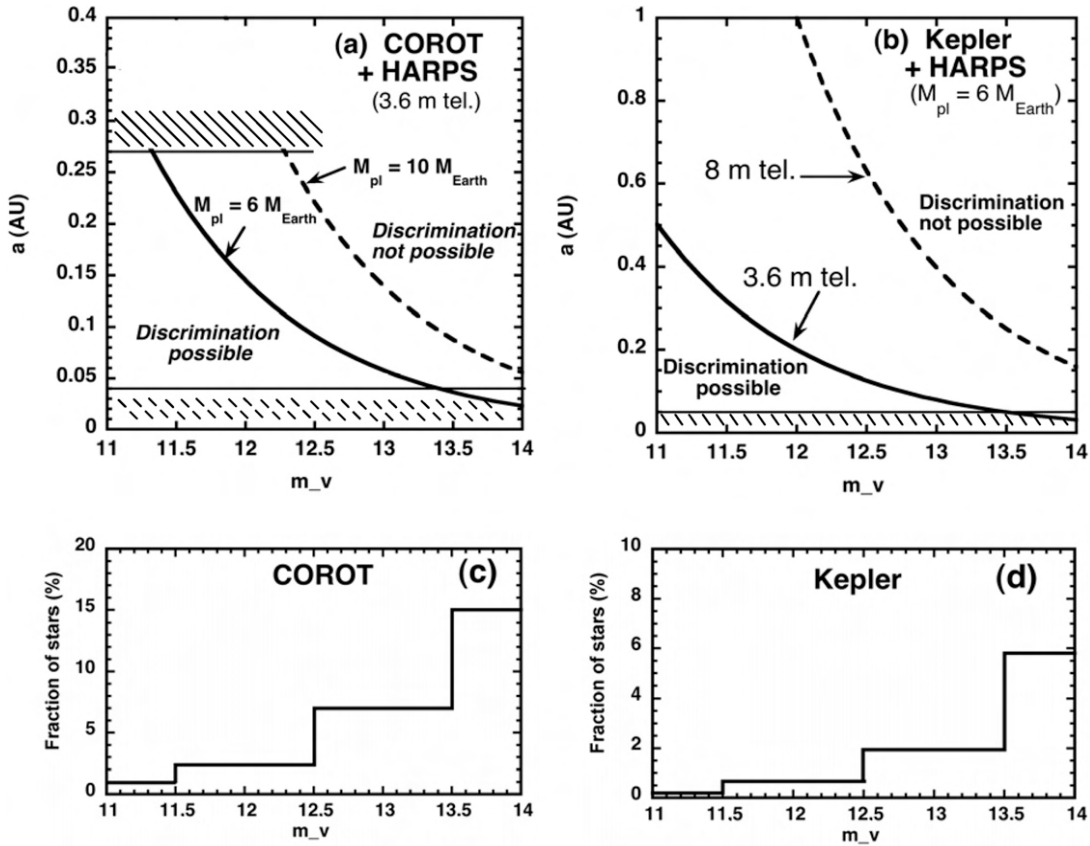

Fig. 6. Regions in the (target magnitude $m_v$, planet to star distance $a$) plane where the discrimination between an OP and a rocky planet can be made at $2\sigma$ (95% confidence) when their masses are $M_{pl} = 6$ or $10\ M_{Earth}$, around GV stars. The corresponding radii of ocean-planets are $R_{OP} = 2.0$, or $2.3\ R_{Earth}$, respectively. The planetary radius is determined using either the CoRoT instrument (a) or the *Kepler* one (b). The mass determination is made with a Doppler velocimeter based on the performances of the HARPS spectrometer mounted on a 3.6 m (40 h of cumulated observations) or, speculatively, on an 8 m telescope. It is assumed that both transit amplitude and stellar reflex velocity accuracies are shot noise-limited. The upper shading indicates a region that is not accessible as a result of instrumental constraints (too long orbital periods). The lower shadings indicate a region where OPs may, or may not, have kept their water, because our estimates of water erosion are only upper limits. Panels (c) and (d) are the histograms of the target stars for CoRoT and *Kepler*, respectively. If the swelling of a hot ocean-planet was confirmed to be of 10% (see Section 2.1), it would significantly help to discriminate between OP and rocky planets. For instance, the boundary for the 6 $M_{Earth}$ planet would approximately move to that of the 10 $M_{Earth}$ case.

Earth-mass or a 10 Earth-mass planet around a GV star can be made at $2\sigma$ (95% confidence). Two transit instruments are considered for the transit measurements, that of the CoRoT and *Kepler* missions. The Doppler determination of the mass is based on the performances of the HARPS spectrometer mounted on a 3.6-m telescope (the present situation) or an 8-m telescope (a speculative situation). It appears that *the target stars of transit missions are faint objects, which is a serious handicap for the Doppler measurements*. As a result, the nature of small planets can be only be determined for the brightest stars ($m_v < 13$) and the shortest-period planets, at least for CoRoT.

This points out the importance of determining how close to their stars OPs are expected to keep their water (Section 5). Our estimates of water erosion are only upper limits. Planets that are inside the boundaries we have found may, or may not, have lost their water. Their exact nature will be constrained by further modeling and determined by the observations.

At fixed magnitude, stars with smaller radii, e.g. K stars, are more favorable, but there are fewer of them than G stars in a magnitude-limited sample.

In the CoRoT target list, the fraction of stars with magnitude 11, 12, 13 and 14 are expected to be 1, 2.3, 7 and 15% of the total number of targets, respectively (Bordé et al., 2003). These stars are not the majority of the target objects, but they are not a minute fraction of them either. However, stars with $m_v \leqslant 12$ will saturate the brightest pixels of the power Spread Function (PSF) in the CCD. Presently, these stars are not considered for photometry because they are technically more difficult to monitor, but this situation may improve sometime after the beginning of the mission. To be optimistic, we will keep these stars in our further estimates, especially as they are those with the highest S/N.

Within the *arbitrary hypothesis* that each star, approximated as a G star (the most probable stellar type in the magnitude-limited CoRoT field), has a 6 $M_E$ [10 $M_E$] planet at 0.05 AU, the number of detections by CoRoT with sufficient accuracy for determining the nature of this planet would be the total number of stars which are bright enough ($m_v \leqslant 13$, according to Fig. 6a) times the probability of transit ($R_{st}/a = 9\%$). For the total mission (5 stellar fields) and $M_{pl} = 6\ M_E$, this number would be the number of stars with $m_v \leqslant 13$, or $5 \times 12,000 \times (1\% + 2.3\% + 7\%)$ times 0.09, or ~550 detections. For $M_{pl} = 10\ M_E$, the limit is $m_v \leqslant 14$, and the corresponding number would be ~1400 detections. In all cases, a long (80 h) Doppler follow-up of the transits on at least a 3.6 m telescope is necessary.

However, *the reader should not misinterpret these numbers*. We are *not* claiming that CoRoT is going to detect hundreds of 10 Earth-mass planets, for the mere reason that it would be impossible to dedicate 40 h of HARPS time per star for such a number of objects (40 h × 550 objects = 22,000 h ~ 3000 nights ~ 8 years full time assuming a 100% efficiency!). The estimate is mainly interesting for the purpose of determining at what level the absence of any detection would limit the abundance of such planets. If the number of planets with masses in the range 6–10 $M_{Earth}$ and distance to their star less than 0.1 AU was similar to the number of hot Jupiters, i.e. 1%, *the number of expected detections would be between 5 and 10*.

Conversely, a non-detection of such planets, if our accuracy estimates are correct, would indicate that they are not present at the level of ~0.5%. If a more accurate Doppler follow-up were possible, those discriminating detections would be more numerous because the mass measurement accuracy is presently the bottleneck (Eq. (28)).

For *Kepler*, according to Fig. 6b, if the follow-up was made with performed with a Doppler velocimeter analogous to HARPS on a 8 m telescope, for $M_{pl} = 6\ M_{Earth}$, one would expect that planets at $a = 0.10$ AU can be characterized around stars with $m_v \leqslant 14.5$. The total number of monitored stars is 100,000. In the *Kepler* field there ~600,000 stars with $m_v \leqslant 16.5$, 15% of which have $m_v \leqslant 14.5$ (D. Koch, personal communication). With the conservative assumption that the distribution in magnitude of the monitored stars is the same as for the other stars of the field, and the assumption of a 1% abundance of hot earths with masses 6–10 $M_{Earth}$, one would expect to be able to characterize 5 to 10 of them. The difference with the CoRoT case seems not to be huge, although it must be pointed out that the prospect of characterizing terrestrial planets with *Kepler* is on a significantly firmer basis because some of these measurements are expected to have a fair S/N, whereas this is not so frequent with CoRoT.

*We point out the key interest of building a Doppler spectrometer with performance similar to that of HARPS, on an 8 m class telescope for the study of terrestrial exoplanets. This is true for the follow-up of CoRoT and it is especially obvious for that of Kepler.*

This is under the optimistic hypothesis that for faint stars ($m_v = 11$–14) both transit amplitude and stellar reflex velocity are basically limited by *shot noise* and not by the stellar variability or instrument noise. If the stellar noise dominates, using large telescopes will not be useful; the only way to (slowly) improve the S/N will be to increase the integration time.

## 4. Discussion: Mass–radius degeneracy between OPs, $H_2$-rich planets and carbon planets?

As pointed out in the introduction, theoretical works (Alibert et al., 2006; Rafikov, 2006) predict the existence of short period planets with a core mass >6 $M_{Earth}$ and a wide diversity of $H_2$–He envelopes: from $M_{env} \ll 1\ M_{Earth}$ (OPs and super-telluric planets), to a few $M_{Earth}$ (like the hot Neptunes in the system HD69830), and finally $M_{env} \gg 1\ M_{Earth}$ (giant planets).

In planetary formation models, the core is initially made of icy–rocky material from beyond the snowline, but it accretes more rocky planetesimals when migrating between the snow line and its final short period orbit. The collisions occurring during the planet formation can lead to a final composition significantly depleted in volatiles compared with a typical cometary composition. Also, the planet can of course consist of refractory materials only, if migration started closer to the snow line and if enough rocky planetesimals are available.

Therefore, a given couple $[R, M]$ may correspond to either an OP of mass $M$ and a negligible $H_2$–He atmosphere or to a planet with $M = M_{core} + M_{env}$. For the same mass, a hydrogen envelope would be significantly more expanded than an $H_2O$ envelope. At the orbital periods relevant for CoRoT, the high temperature due to the strong irradiation would produce an even more important swelling of the $H_2$–He envelope. An $H_2$–He envelope more massive than about 0.1 $M_{Earth}$ would produce a planetary radius of more than 3 $R_{Earth}$, which is larger than an 10 $M_{Earth}$ OP. To give an example (kindly provided by I. Baraffe), a planet at 0.1 AU, with a rocky core of [$M_{core} = 7.5\ M_{Earth}$, $R_{core} = 1.76\ R_{Earth}$] and an $H_2$–He envelope of 2.5 $M_{Earth}$, would have a total radius of 5.6 $R_{Earth}$. Therefore, to mimic an OP with no $H_2$–He, the mass of $H_2$–He surrounding a rocky core has to be significantly less massive than the water reservoir of this OP. Observing a planet with this precise amount of $H_2$–He, obtained by accretion of the gaseous envelope and its subsequent partial loss by atmospheric escape, appears qualitatively rather unlikely.

Although the diversity of mass–radius relationships should clearly be addressed quantitatively using a model of planet formation (including migration), thereby providing a realistic distribution of $M$ ($M_{Rocks}$, $M_{Ice}$, $M_{H_2}$) versus orbital distance, we conclude here that the identification of massive OPs cannot rigorously be done with a single case. To show up among transiting planets having a wide range of $H_2$–He envelope masses, OPs with negligible $H_2$–He atmospheres have to be relatively abundant. Regarding this point, *short-period, low-mass OPs* ($M < 6\ M_{Earth}$, $a < 0.1$ AU) *provide the best case*, because the accretion of gas is not expected to be important for these masses (Rafikov, 2006), while the escape of $H_2$ is fast (much faster than $H_2O$) at these orbital distances (Baraffe et al., 2005). More massive, $H_2$-deficient OPs at longer periods may be difficult to distinguish within a population of planets (OPs or telluric) with $H_2$ envelopes.

It is also important to note that all ratios $M_{Ice}/M_{Rocks}$ between ~0 and 1 are expected. For a given planetary mass, planetary radii should be found at any value between that of a silicate planet and our OP prototypes (i.e. between the two curves of Fig. 1). For $M > 6$ $M_{Earth}$ we can expect a two distributions of objects, hopefully separated by a gap, for the OP prototypes and the hot Neptune/Saturn/Jupiter objects.

Kuchner and Seager (2005) proposed the existence of carbon planets, formed in environments characterized by C/O > 1. In the absence of detailed modeling of their structure and of their radius as a function of their mass, it is difficult to know if these carbon planets could be identified or confused with silicate-dominated planets or OPs. That is an open question for further studies. What we can say at this point is that the C/O ratio can be measured (within some significant uncertainties) in stellar photospheres, so we should be aware, for *the rare (any?) main sequence stars with high C/O ratio* where a transit will be observed, that a carbon planet might be there. Even for stars with a "solar" C/O ratio, planets (e.g. Jupiter) can still be significantly enriched in carbon. In addition, the icy material found beyond the snow line does not consists only of $H_2O$ ice, $CO_2$, $CH_4$ and CO are also important constituents. The accretion of one Earth mass of $H_2O$ ice implies the accretion of about 0.1 Earth mass of these carbon-based volatile, which can have a significant effect on the structure and the evolution of the planet. We should thus remain aware that a large diversity of elemental composition can exist among planets and that cannot be characterized only by the $w = water/silicate$ ratio. Nevertheless, the detection by CoRoT or Kepler, and around K,G or F stars, of short-period low-mass planets ($T_{orb} \leqslant 1$ month, $M_{pl} \leqslant 6$ $M_{Earth}$), with radii significantly larger than silicate-dominated planets would unveil a population of volatile-rich planets. The best candidate for this volatile being $H_2O$.

## 5. Conclusion

We have revisited the theoretical estimate of the physical density of hot ocean-planets (OPs) and concluded that the results established for planets in the habitable zone (HZ) are still valid much closer to their star, because their expected swelling should remain limited when compared to the planetary radius. Curiously enough, *a very hot OP has a thick water atmosphere in direct contact with a (high pressure) ice mantle*.

Models of the internal structure of OPs yield a mean density smaller than that of rocky planets. For the same mass there is a difference in radius of ∼20%, that can reach 30% for the hottest OPs, due to the swelling of their water envelope In this paper we have considered whether an identification of OPs based on the measurement of their densities is actually possible in the near- to mid-future.

If strong enough, migration could bring these speculative planets to within the close vicinity of their parent star, where their masses and radii can be better measured by Radial Velocity and by transit photometry. CoRoT will detect planets with periods less than 75 days, whereas *Kepler* will have access to periods less than 1.33 years. In both cases, *close-in planets will be the easiest ones to detect and to characterize*.

We have estimated the erosion rate of a hot ocean-planet by thermal escape driven by exosphere heating (with X and UV radiation) and by non-thermal escape driven by ejected particles from the star. We found that the water reservoir is only weakly affected, so that an OP should keep its peculiar composition even fairly close to its star. For example, in front of a solar-type star, a 6 Earth-mass OP should retain most of its water content for more than 5 Gyrs if it is beyond 0.04 AU from its star.

Then, we have addressed the accuracy needed to discriminate between OPs and rocky planets, assuming that the planetary models are adequate. The sources of uncertainty on the planetary density are those on the mass determination by Radial Velocity measurements, the stellar radius determination, and the photometric measurement during the transits. As expected, the accuracy of the *Kepler* photometry is higher than that of CoRoT. However with the presently available instruments, *it is the uncertainty on Radial Velocity measurements that is the limiting factor for expected detections by both transit missions*. As a result, the determination of the nature of these planets seems possible only in the adequate domains of the ($m_v$, a) plane that are shown in Fig. 6 for both missions. They correspond to the *brightest stars* ($m_v < 14$) and to *the planets closest to their star*, especially for CoRoT, but beyond the limit for the survival of their water reservoir.

If each star had a 6–10 Earth mass planet at ∼0.10 AU, an arbitrary hypothesis, the number of detections with CoRoT would be several hundred. Conversely, the absence of detection of such planets would indicate that they are not present at the level of ∼1%.

A clear conclusion of our study is *that full benefit of the high photometric precision of CoRot, and particularly* Kepler, *can be obtained only if a new generation of Radial Velocity instruments is built* that can make accurate measurements on faint stars. In that case, the identification of OPs could be done on a significantly larger sample of stars. In the meantime, a significant fraction of high performance Doppler velocimeters such as HARPS should be devoted to the follow-up of the most interesting candidates found by CoRoT, if we hope to be able to discriminate between the different possible compositions of terrestrial planets.

### Acknowledgment

We are grateful to Odile Dutuit, Roland Thissen, Alexandro Morbidelli, and David Koch for valuable discussions on the water molecule absorption in the EUV, the possible abundance of these putative planets and precise information on *Kepler*, respectively. We are also grateful to an anonymous referee and to Marc Kuchner for valuable comments on our initial manuscript. This work was supported by CNRS and CNES in part by the Programme National de Planétologie and the Groupe de Recherche "Exobio," and under contract 1256791 with the Jet Propulsion Laboratory (JPL) funded by NASA through the Michelson Fellowship Program. We also acknowledge the benefit from the ISSI Team "Evolution of Habitable Planets."